\documentclass[twocolumn, 11pt]{article}
\newcommand{\shorttitle}{Portfolio-based Constrained-Multi-Objective Bayesian Optimization for Materials Design}
\usepackage{pdfpages}
\usepackage{draftstyle}
\usetikzlibrary{positioning,fit,arrows.meta}

\newcommand{\x}{\mathbf{x}}
\newcommand{\X}{\mathcal{X}}
\newcommand{\fvec}{\mathbf{f}}
\newcommand{\cvec}{\mathbf{c}}
\newcommand{\HV}{\mathrm{HV}}

\newcommand{\R}{\mathbb{R}}

\newcommand{\A}{\mathcal{A}}
\newcommand{\D}{\mathcal{D}}
\newcommand{\F}{\mathcal{F}}
\newcommand{\Pset}{\mathcal{P}}

\title{Portfolio-Based Constrained Multi-Objective Bayesian Optimization for Materials Design}

\author{
  Sushant Sinha$^{1*}$, Christofer Hardcastle$^{1}$, Robert Robinson$^{1}$, Shakti Prasad Padhy$^{2}$, \\ Brent Vela$^{1}$, Douglas Allaire$^{2}$,  and Raymundo Arróyave$^{1,2,3}$
}

\date{
\footnotesize
  $^{1}$Department of Materials Science and Engineering, Texas A\&M University, College Station, TX, USA\\
  $^{2}$J. Mike Walker '66 Department of Mechanical Engineering, Texas A\&M University, College Station, TX, USA\\
  $^{3}$Wm Michael Barnes '64 Department of Industrial and Systems Engineering, Texas A\&M University, College Station, TX, USA\\
  $^{*}$Corresponding author: sushant.sinha@tamu.edu
}

\begin{document}
\maketitle

\begin{abstract}
 
  Materials discovery and design campaigns can be formulated as constrained multi-objective Bayesian optimization (CMOBO) problems, within which each experimental decision negotiates between two coupled but competing goals: discovering feasible candidates and refining the underlying Pareto front. Here we recast acquisition-function choice as an adaptive policy-selection problem over a portfolio of conventional and feasibility-focused acquisition functions. This was done using two controllers: UCB-Bandit, a modified UCB multi-armed bandit, and Agentic-Switch, a multi-agent decision system driven by a large language model (LLM). Both were evaluated against fixed-policy baselines \textit{in silico} across five synthetic benchmark functions and two materials design case studies. The adaptive policies performed competitively in terms of both cumulative feasibility count and feasible hypervolume improvement, while each individual acquisition function performed well for only one metric, suggesting that adaptive policies are better suited for constrained materials science problems.
\end{abstract}

\keywords{Materials Design; AI4Materials; Constraint Satisfaction; Bayesian Optimization; Adaptive Policy; LLM-Agents; Bandit Algorithms}

\section{Introduction}
\label{sec:introduction}

Sample efficiency is critical in materials discovery and design campaigns, as each iteration corresponds to a costly experiment or computational simulation. Because practical campaigns are typically multi-objective and constrained, the task becomes one of identifying candidates that are both feasible and non-dominated along a multidimensional Pareto frontier \cite{macleod2022self, low2022mapping, suzuki2020multi, khatamsaz2022multi, khatamsaz2023bayesian}. Traditionally, these problems have been framed as optimization problems aimed at identifying Pareto-optimal solutions. However, \emph{Pareto-optimality alone} does not guarantee satisfaction of the strict feasibility constraints required for real-world problems \cite{feliot2017bayesian, khatamsaz2022multi, khatamsaz2023bayesian, maguire2025good}. Constraint satisfaction, a design framework that focuses on satisfying all predefined requirements (regardless of conventional optimality), has been proposed as an alternative design strategy in materials science \cite{maguire2025good,Arroyave2016,Galvan2017,AbuOdeh2018,Vela2023,Sohst2022}. Yet, emphasizing feasibility alone can sacrifice Pareto quality or hypervolume growth, since constraint-satisfying candidates need not be non-dominated \cite{maguire2025good}.

Constrained multi-objective Bayesian optimization (CMOBO) provides a natural middle ground between these two formulations. Because the feasible region and the feasible Pareto frontier must both be learned under the same limited budget, each query negotiates between discovering feasible regions and improving the hypervolume of the feasible non-dominated set \cite{feliot2017bayesian,khatamsaz2022multi,khatamsaz2023bayesian,garrido2019predictive}. 
These two goals do more than compete for evaluations, since progress on one conditions what the other can achieve. Hypervolume cannot grow until feasible points have been found. Conversely, if the feasible region of the underlying response surface has been sufficiently learned, then focusing on feasibility sacrifices hypervolume improvement.
 The consequences of this coupling are visible even in simple combinations of acquisition functions. Maguire et al. \cite{maguire2025good} found that the product of expected hypervolume improvement (EHVI) \cite{daulton2020differentiable} and probability of feasibility (PoF) \cite{gardner2014bayesian} performed intermediately with regards to both hypervolume improvement and cumulative feasibility query count, with pure EHVI resulting in greater hypervolume improvement and pure PoF finding more feasible queries. Neither strategy was fully exploited, even though the pairing was meant to serve both.

Alternatively, Khatamsaz et al. \cite{khatamsaz2022multi,khatamsaz2023bayesian} demonstrated that constrained materials design can be treated as a coupled Bayesian classification and multi-objective Bayesian optimization problem. Gaussian-process classifiers there actively learn the feasibility boundaries, while Gaussian-process regressors and EHVI refine the feasible Pareto frontier. Such feasibility-first workflows still spend a substantial part of the campaign on constraint-boundary learning before Pareto refinement, which suits large \textit{in silico} design spaces but is less attractive on a small experimental budget. COMBOO \cite{li2025combo} resolves the tension differently, confining the search to an optimistic feasible region estimated from constraint upper confidence bounds. Random scalarization of the objective bounds inside that region then drives refinement, with cumulative bounds on hypervolume regret and on constraint violation. However, this approach commits to a single rule for the whole campaign, which leaves open whether the balance between feasible discovery and feasible-hypervolume improvement is stable enough for a fixed rule to serve.  
 
The broader BO literature has long recognized that acquisition-function choice is problem dependent \cite{brochu2011portfolioallocationbayesianoptimization, shahriari2015entropysearchportfoliobayesian, ngo2026adaptiveacquisitionselectionbayesian}. Brochu et al., for example, proposed hedging approaches that treat each acquisition function as an arm in a multi-armed bandit problem \cite{brochu2011portfolioallocationbayesianoptimization}. Their best model, GP-Hedge, outperformed the strongest individual acquisition function on each test function. Shahriari et al. achieved similar results with their Entropy Search Portfolio (ESP), an information-theoretic portfolio criterion \cite{shahriari2015entropysearchportfoliobayesian}. Sheikh and Marcus extended portfolio allocation to multi-objective problems with HedgeMO, which normalizes rewards across acquisition functions so that a single hedging rule can operate on them \cite{sheikh2022bayesian}.

These methods all select among acquisition functions, yet the selection does not respond to how the campaign is progressing. Ngo et al. address this with LMABO, an adaptive acquisition-function selection method in which an LLM agent with access to campaign information acts as the controller \cite{ngo2026adaptiveacquisitionselectionbayesian}. None of this work, however, addresses the decision faced here, where feasible discovery and feasible-hypervolume improvement compete for the same expensive evaluations.

We extend this decision-theoretic approach with two controllers that switch between acquisition strategies drawn from a shared portfolio: Agentic-Switch, a multi-agent adaptive policy selector, and UCB-Bandit, a custom upper-confidence-bound multi-armed bandit. Both were evaluated against the individual acquisition strategies on five constrained multi-objective benchmarks, namely Constrained Branin-Currin, CONSTR, two- and three-objective C2-DTLZ2, and Disc Brake. Two \textit{in silico} materials design case studies follow, one on thermoelectric materials and one on a high entropy alloy (HEA) system. Across these problems, the adaptive controllers gave the best balance between feasibility seeking and hypervolume improvement, while remaining competitive on each individually. Our results demonstrate that adaptive policy selection provides a robust, problem-agnostic strategy for CMOBO, attaining competitive performance on both feasibility and hypervolume metrics across problems with widely varying feasibility regimes, i.e., from moderately constrained synthetic benchmarks to severely sparse real-world materials design pools.

\section{Results}
\label{sec:results}
\subsection*{Benchmark problems}

\begin{figure*}
    \centering
    \includegraphics[width=1.0\linewidth]{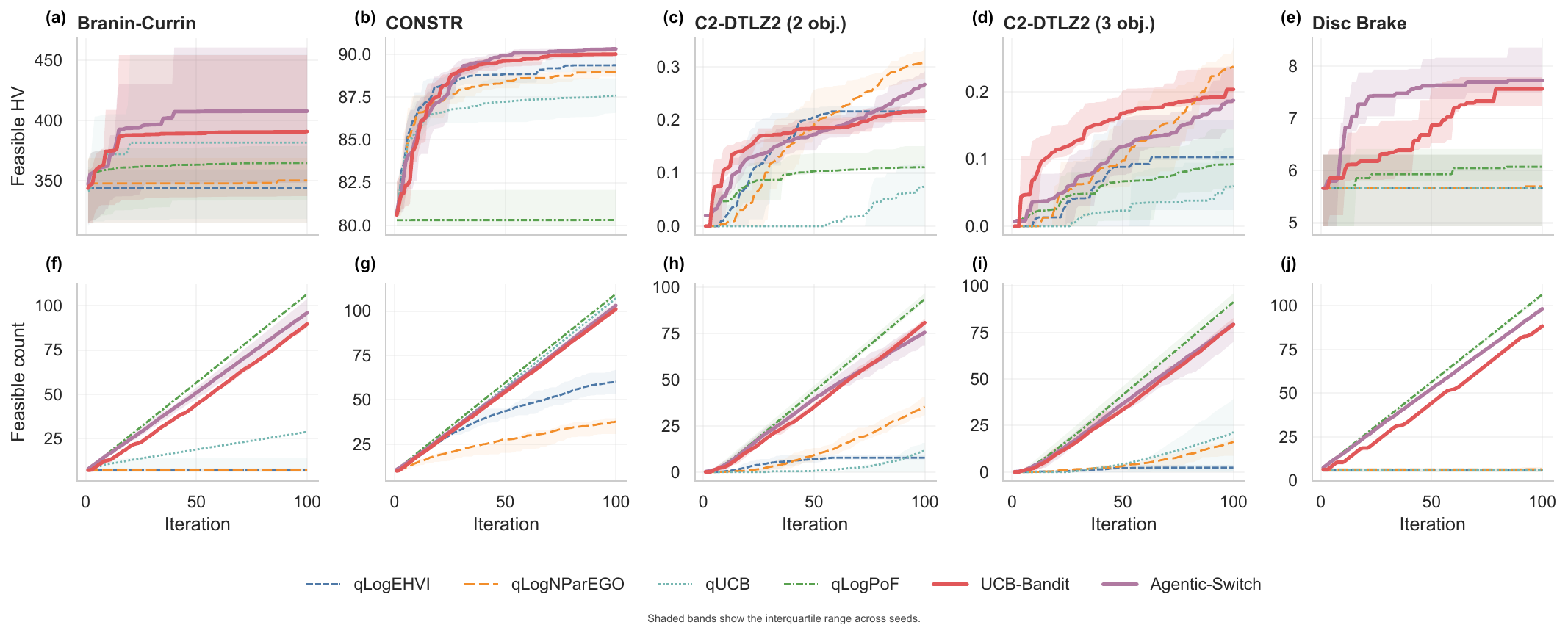}
    \caption{\textbf{Performance trajectories on constrained multi-objective benchmark functions.}
    Five synthetic benchmarks are shown with feasible hypervolume in the top row and cumulative feasible discoveries in the bottom row. Curves report the mean over 10 random seeds, and shaded bands indicate the interquartile range across seeds at each iteration. Adaptive policies are competitive across several benchmarks, while fixed policies reveal distinct specialization: \texttt{qLogPoF} often maximizes feasible discoveries, whereas \texttt{qLogNParEGO} is strongest for feasible hypervolume on the C2-DTLZ2 problems.}
    \label{fig:benchmark-trajectories}
\end{figure*}

We evaluated the fixed-acquisition policies and the portfolio-based adaptive methods on five constrained multi-objective benchmark functions: constrained Branin--Currin, CONSTR, two- and three-objective C2-DTLZ2 variants, and Disc Brake. Figure \ref{fig:benchmark-trajectories} shows that \texttt{qLogPoF} produced the largest feasible counts on every benchmark, though these did not translate into the largest feasible hypervolume. CONSTR is the clearest case, where \texttt{qLogPoF} obtained the largest feasible count while remaining weakest in feasible hypervolume. Disc Brake followed the same pattern, with extra feasible points easy to accumulate but most of them leaving the feasible frontier where it was. In contrast, \texttt{qLogNParEGO} achieved the highest final feasible hypervolume on the two C2-DTLZ2 problems, at considerably lower feasible counts than \texttt{qLogPoF}. That combination points to a geometry in which sustained Pareto refinement was worth more than continued harvesting of feasible points. 

These problems serve as useful boundary conditions, since the fixed acquisition policies expose the underlying conflict in constrained MOBO. The policy that finds feasible candidates efficiently need not be the policy that attains the best feasible Pareto hypervolume, and the reverse holds as well. Adaptive switching, however, performed better on both metrics. Agentic-Switch and UCB-Bandit were each competitive across the suite in mean feasible hypervolume, with Agentic-Switch highest on Branin--Currin, CONSTR, and Disc Brake. Both adaptive policies also stayed close to \texttt{qLogPoF} in mean feasible count, the axis on which the strongest hypervolume policies performed worst.

We then turned to two materials design case studies, one on thermoelectric (TE) materials and one on high entropy alloy (HEA) design. Both drew on finite candidate pools, which sharpened the feasibility bottleneck well beyond what the synthetic benchmarks impose and brought it closer to what a real campaign encounters.

\subsection*{Thermoelectric Materials Design}

The TE study presented a more constrained feasible region than the benchmarks, though one that remained learnable, with a pool of 3,138 candidates of which 144 were feasible and 82 were both feasible and non-dominated. A fixed \texttt{qLogPoF} policy was a strong baseline here, since feasibility information stayed useful throughout the campaign. Agentic-Switch nonetheless reached a higher feasible count while ending with slightly higher feasible hypervolume (figure \ref{fig:materials-trajectories}). It held to \texttt{qLogPoF} for most iterations (\(\sim 81\%\)), turning to \texttt{qUCB} (\(\sim 14\%\)) or \texttt{qLogNParEGO} (\(\sim 5\%\)) when recent \texttt{qLogPoF} selections had stalled. Figure \ref{fig:agentic-decision-trace} shows one such exchange among the three agents, a pattern that amounts to feasibility-dominant control with an escape route when feasibility progress stalls locally. UCB-Bandit recovered slightly fewer feasible candidates than fixed \texttt{qLogPoF}, yet finished with a feasible hypervolume improvement comparable to Agentic-Switch. It reached that result through a broader mixture: \texttt{qLogPoF} (\(\sim 47\%\)), \texttt{qUCB} (\(\sim 23\%\)), \texttt{qLogEHVI} (\(\sim 15\%\)), and \texttt{qLogNParEGO} (\(\sim 14\%\)). Every remaining fixed acquisition strategy performed worse on both metrics.

\subsection*{High Entropy Alloy Design}

The HEA study posed a far more severe feasibility bottleneck, with only 18 of 40,553 candidate alloys both feasible and Pareto optimal. We adopted it to test the approach in a more realistic setting, one carrying hard constraints such as manufacturability that further restrict which candidates can be measured at all. The probability of forming a single BCC phase (\(p_{\mathrm{BCC}}\)) stood in for such a constraint here.

We used the same data and problem definition as Maguire et al. \cite{maguire2025good}, weighting every arm in the portfolio by \(p_{\mathrm{BCC}}\) rather than adding \(p_{\mathrm{BCC}}\) as a separate arm, which would have required a methodological change. We also adopted their \texttt{pEHVI} and \texttt{PoF} formulations, which replaced \texttt{qLogEHVI} and \texttt{qLogPoF} in the portfolio. Even so, the acquisition balance remained precarious. A hypervolume-focused policy could spend many measurements without satisfying the constraints on the objectives, while a purely feasibility-driven one could settle into a narrow region and leave the feasible frontier where it was.

\begin{figure*}
    \centering
    \includegraphics[width=0.75\linewidth]{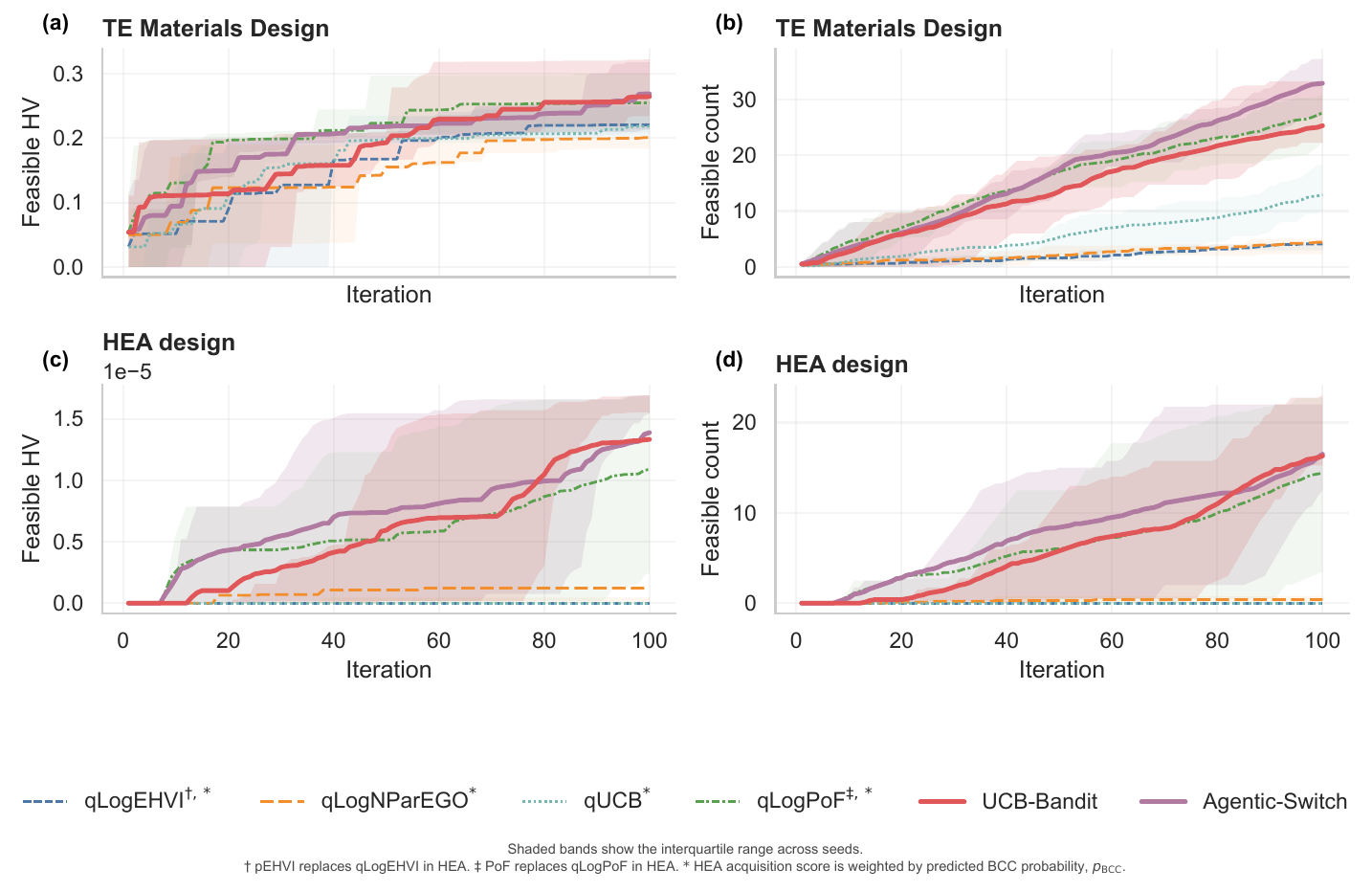}
    \caption{\textbf{Adaptive acquisition control on finite-pool materials design studies.}
    Feasible hypervolume and cumulative feasible discoveries are shown for the thermoelectric material and HEA design tasks. Curves report the mean over 10 paired random seeds, with interquartile bands across seeds. Adaptive policies are especially competitive in the materials studies, where feasible candidates are sparse and feasible discovery must be balanced against Pareto-front improvement. In the HEA study, \(^{\dagger}\)pEHVI replaced \texttt{qLogEHVI}, \(^{\ddagger}\)PoF replaced \texttt{qLogPoF}, and \(^{*}\)HEA acquisition scores were weighted by predicted BCC probability, \(p_{\mathrm{BCC}}\).}
    \label{fig:materials-trajectories}
\end{figure*}

The fixed policies showed exactly that failure mode (figure \ref{fig:materials-trajectories}). \texttt{PoF} was the only fixed acquisition to recover a substantial number of feasible alloys from the pool. \texttt{pEHVI} and \texttt{qUCB} consumed many full measurements without returning a single alloy that satisfied the constraints on the objectives. Both adaptive policies improved modestly on fixed \texttt{PoF}, though they arrived there by different routes. Agentic-Switch selected \texttt{PoF} in $\sim 76\%$ of HEA decisions and \texttt{pEHVI} in $\sim 18\%$, using \texttt{qLogNParEGO} only sparsely. UCB-Bandit spread its choices more evenly, selecting \texttt{PoF} in $\sim 47\%$ of decisions and \texttt{qLogNParEGO}, \texttt{pEHVI}, and \texttt{qUCB} in roughly $18\%$ each. That higher switching rate matched Agentic-Switch in feasible count but bought no advantage in feasible hypervolume. The adaptive policies drew on the non-feasibility-seeking arms in different proportions, yet those arms performed substantially worse as fixed policies on both feasibility and hypervolume. Limited, state-dependent use of such arms therefore appears to complement a feasibility-dominant strategy, as it did in the TE problem.

\begin{figure*}
    \centering
    \definecolor{agentstate}{HTML}{F6F7F9}
    \definecolor{agentfeas}{HTML}{EAF3F8}
    \definecolor{agentfeasline}{HTML}{2F6F91}
    \definecolor{agenthv}{HTML}{FFF3E4}
    \definecolor{agenthvline}{HTML}{C87519}
    \definecolor{agentarb}{HTML}{F2EEF7}
    \definecolor{agentarbline}{HTML}{7A5B9E}
    \begin{tcolorbox}[
        enhanced,
        colback=white,
        colframe=black!18,
        boxrule=0.6pt,
        arc=1.4mm,
        left=0.75em,
        right=0.75em,
        top=0.75em,
        bottom=0.75em,
        title=\textbf{Representative Agentic-Switch decision trace},
        coltitle=darkgray,
        attach boxed title to top left={xshift=0.8em,yshift=-0.35em},
        boxed title style={colback=white,boxrule=0pt,arc=1mm}
    ]
    \begin{tcolorbox}[
        enhanced,
        colback=agentstate,
        colframe=agentstate,
        boxrule=0pt,
        arc=1mm,
        left=0.7em,
        right=0.7em,
        top=0.45em,
        bottom=0.45em
    ]
    \footnotesize
    \textbf{Logged iteration.} ESTM thermoelectric study, seed 5, iteration 81.
    \hfill
    \textbf{Selected acquisition:} \texttt{qucb}
    \hfill \\
    \textbf{Observed feasible:} 1
    \hfill
    \textbf{Feasible-HV gain:} 0.0048137939175743
    \end{tcolorbox}

    \vspace{0.55em}

    \begin{tcolorbox}[
        enhanced,
        colback=agentfeas,
        colframe=agentfeasline,
        boxrule=0.45pt,
        arc=1mm,
        left=0.7em,
        right=0.7em,
        top=0.55em,
        bottom=0.55em,
        title=\textbf{Feasibility advocate},
        coltitle=black,
        boxed title style={colback=agentfeasline!12,boxrule=0pt,arc=0.8mm},
        attach boxed title to top left={xshift=0.5em,yshift=-0.3em}
    ]
    \scriptsize
    \begin{tabularx}{\linewidth}{@{}>{\ttfamily}p{0.15\linewidth}>{\raggedright\arraybackslash}X@{}}
    recommended\_acq & qlogpof\\
    reason & Protect feasible discovery\\
    argument & The recent history shows that 'qucb' has been successful in finding feasible points, but its predicted feasibility is extremely low, indicating that its recent success might be due to chance rather than a reliable strategy. 'qlogpof' has the highest predicted feasibility among the arms and has been consistently ranked highest for feasibility and hypervolume proxy, despite not achieving recent feasible gains. Given the primary goal of protecting feasible discovery, 'qlogpof' is the best choice to ensure a more reliable path to feasible solutions.\\
    confidence & high
    \end{tabularx}
    \end{tcolorbox}

    \vspace{0.25em}
    \begin{tcolorbox}[
        enhanced,
        colback=agenthv,
        colframe=agenthvline,
        boxrule=0.45pt,
        arc=1mm,
        left=0.7em,
        right=0.7em,
        top=0.55em,
        bottom=0.55em,
        title=\textbf{Hypervolume advocate},
        coltitle=black,
        boxed title style={colback=agenthvline!14,boxrule=0pt,arc=0.8mm},
        attach boxed title to top left={xshift=0.5em,yshift=-0.3em}
    ]
    \scriptsize
    \begin{tabularx}{\linewidth}{@{}>{\ttfamily}p{0.15\linewidth}>{\raggedright\arraybackslash}X@{}}
    recommended\_acq & qucb\\
    reason & Maximize feasible hypervolume growth\\
    argument & Despite 'qlogpof' having the highest predicted feasibility, it has not resulted in any feasible gains recently. 'qucb', on the other hand, has achieved feasible gains in the last two iterations, contributing to hypervolume growth. The recent success of 'qucb' suggests it is currently more effective at discovering feasible solutions that contribute to hypervolume, even if its predicted feasibility is low. Given the recent history and the need to maximize feasible hypervolume growth, 'qucb' is the best choice.\\
    confidence & 0.6
    \end{tabularx}
    \end{tcolorbox}

    \vspace{0.25em}
    \begin{tcolorbox}[
        enhanced,
        colback=agentarb,
        colframe=agentarbline,
        boxrule=0.45pt,
        arc=1mm,
        left=0.7em,
        right=0.7em,
        top=0.55em,
        bottom=0.55em,
        title=\textbf{Arbiter},
        coltitle=black,
        boxed title style={colback=agentarbline!14,boxrule=0pt,arc=0.8mm},
        attach boxed title to top left={xshift=0.5em,yshift=-0.3em}
    ]
    \scriptsize
    \begin{tabularx}{\linewidth}{@{}>{\ttfamily}p{0.15\linewidth}>{\raggedright\arraybackslash}X@{}}
    acq & qucb\\
    reason & Maximize feasible hypervolume growth\\
    reflection & Despite the low predicted feasibility of 'qucb', its recent performance has led to feasible discoveries and hypervolume gains, indicating a potential for continued success in the current context. The primary goal of maximizing feasible hypervolume over the run aligns with the recent effective performance of 'qucb'. While 'qlogpof' has a higher predicted feasibility, its lack of recent feasible gains suggests that 'qucb' is currently the more effective choice for achieving the run-level objective.\\
    confidence & 0.65
    \end{tabularx}
    \end{tcolorbox}
    \end{tcolorbox}
    \caption{\textbf{Example of the three-agent decision structure.}
    Exact logged feasibility-advocate, hypervolume-advocate, and arbiter fields are shown for one Agentic-Switch iteration in the  thermoelectric materials design study. The example shows how the controller can move away from the arm with the highest predicted feasibility when recent outcomes indicate that another arm is producing feasible-hypervolume gains.}
    \label{fig:agentic-decision-trace}
\end{figure*}

\subsection*{Feasibility-Hypervolume State Space}

\begin{figure*}
    \centering
    \includegraphics[width=1\linewidth]{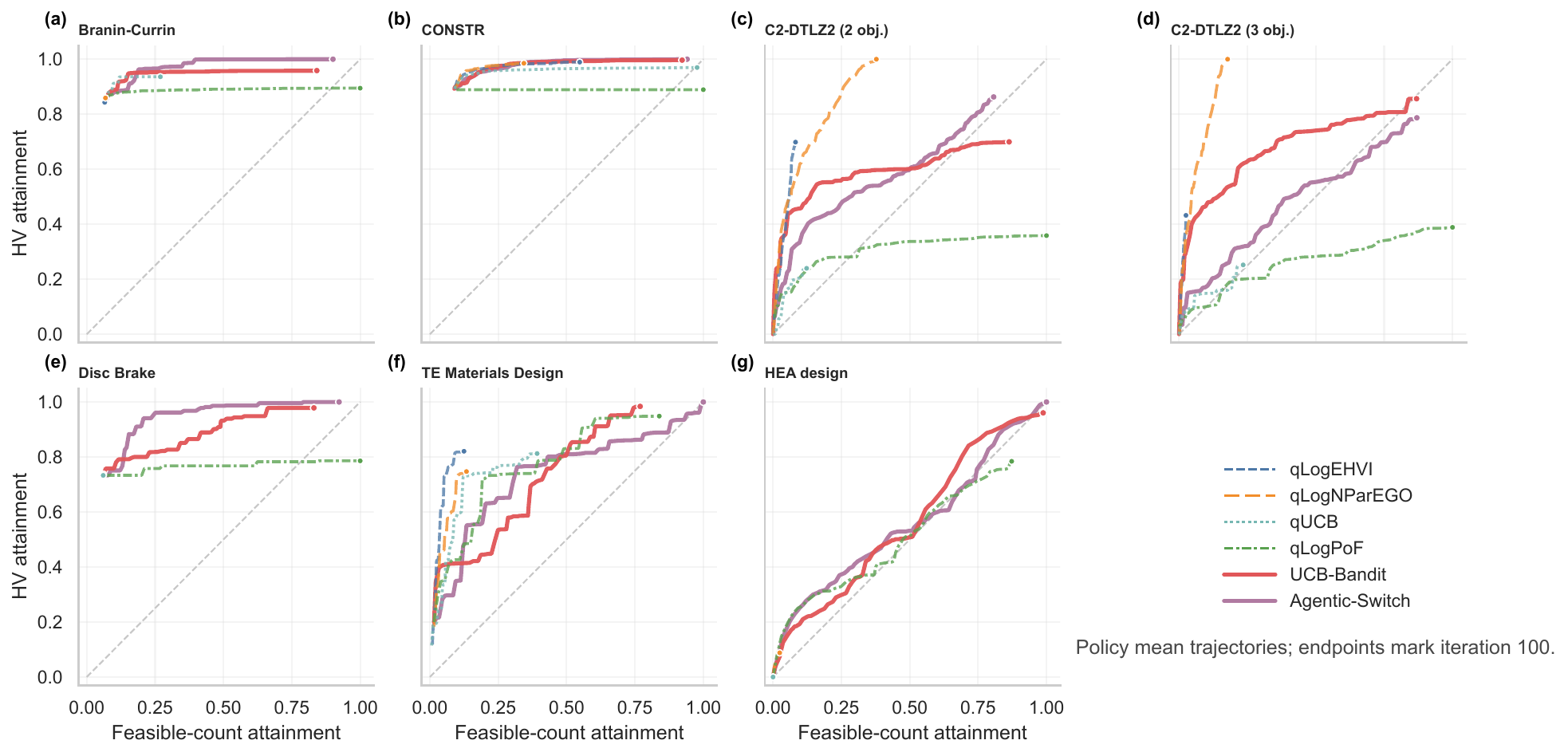}
    \caption{\textbf{Policy trajectories in feasibility-HV state space.}
    Each panel shows policy mean trajectories over 100 BO iterations, with the x-axis giving feasible-count attainment and the y-axis giving feasible-hypervolume attainment. Both axes were normalized within each problem by the best final value observed across policies. Endpoints mark iteration 100. The plot summarizes how policies convert budget into constrained search progress: fixed policies often specialize toward either feasible discovery or hypervolume improvement, whereas adaptive policies frequently trace more balanced paths through the feasible-count and feasible-HV state space.}
    \label{fig:state-space-trajectories}
\end{figure*}

Figure \ref{fig:state-space-trajectories} maps every policy trajectory into a normalized feasibility-hypervolume state space, with feasible-count attainment on the horizontal axis and feasible-hypervolume attainment on the vertical. The question this view poses is whether a policy is only accumulating feasible candidates or is converting those discoveries into Pareto-frontier improvement. The fixed-acquisition policies traced characteristic directions through the space. Feasibility-driven policies moved rapidly to the right, which indicates efficient discovery, but their hypervolume progress stalled when the additional feasible points were dominated or lay far from the feasible frontier. Pareto-oriented policies moved upward efficiently where feasible candidates were already at hand, and stagnated where feasibility was sparse. The adaptive policies instead followed a diagonal path, converting feasible discovery into frontier improvement as the campaign went on. \textit{This geometric perspective crystallizes the central result of this work.}

\subsection*{Behavioral analysis of adaptive policies}

\begin{figure*}
    \centering
    \includegraphics[width=0.75\linewidth]{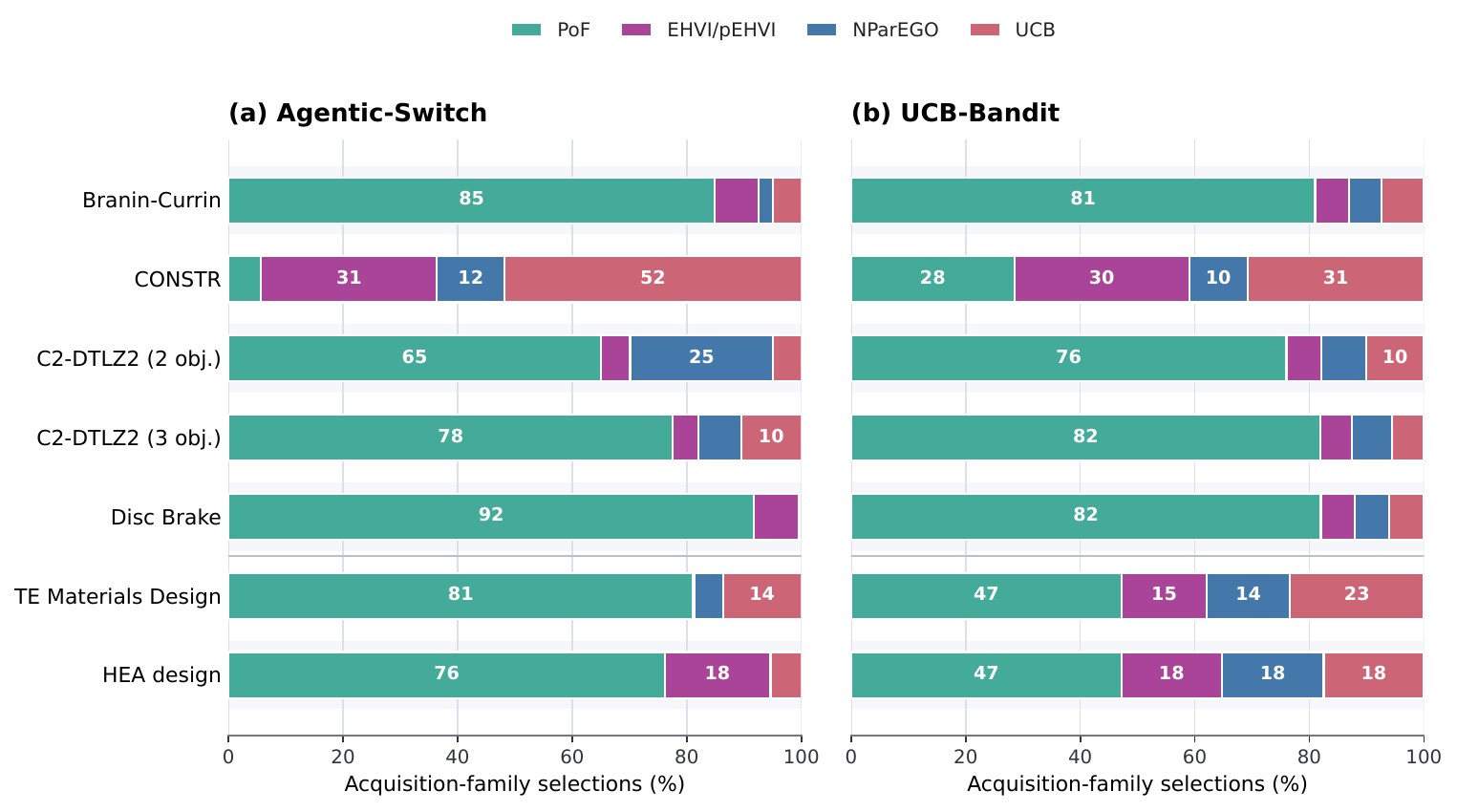}
    \caption{\textbf{Adaptive acquisition-family usage across problems.}
    Stacked bars show the fraction of optimization iterations assigned to each acquisition family by Agentic-Switch and UCB-Bandit. Percentages are computed across 10 seeds and 100 iterations per problem for each adaptive policy. Fixed policies were omitted because their acquisition choice is deterministic. Agentic-Switch is generally more feasibility-anchored, with high PoF-family usage on most problems, whereas UCB-Bandit mixed acquisition families more broadly, especially on CONSTR, TE Materials Design, and HEA design. In the HEA study, pEHVI was grouped with the EHVI family and PoF replaced \texttt{qLogPoF}.}
    \label{fig:main-adaptive-acquisition-usage}

\end{figure*}

The two adaptive policies differed in how they used the acquisition portfolio, and we quantified that difference with a post-hoc analysis of the completed runs. Figure \ref{fig:main-adaptive-acquisition-usage} gives the average share of each arm for both controllers on every problem, averaged over seeds. Agentic-Switch was feasibility anchored almost everywhere, with the PoF arm taking between $65\%$ and $92\%$ of its decisions on six of the seven problems. CONSTR was the exception, where \texttt{qUCB} became the dominant choice. That reversal is consistent with the trajectories in figure \ref{fig:benchmark-trajectories}(b) and \ref{fig:benchmark-trajectories}(g), where a fixed feasibility-seeking policy accumulated feasible points without converting them into feasible-hypervolume progress.

UCB-Bandit showed a different signature, spreading its selections more broadly on CONSTR and on the two materials studies. Its scalar reward was dominated by feasibility, and many selected actions returned no reward at all. Mean selected reward and its feasibility component both shifted across the early, middle, and late stages of a campaign, so the signal the controller learned from was non-stationary. On Branin-Currin, C2-DTLZ2, and Disc Brake, \texttt{qLogPoF} drew a consistent reward and UCB-Bandit stayed PoF-dominant, with occasional sampling of other arms attributable to the UCB bonus.

\begin{figure*}
    \centering
    \includegraphics[width=1\linewidth]{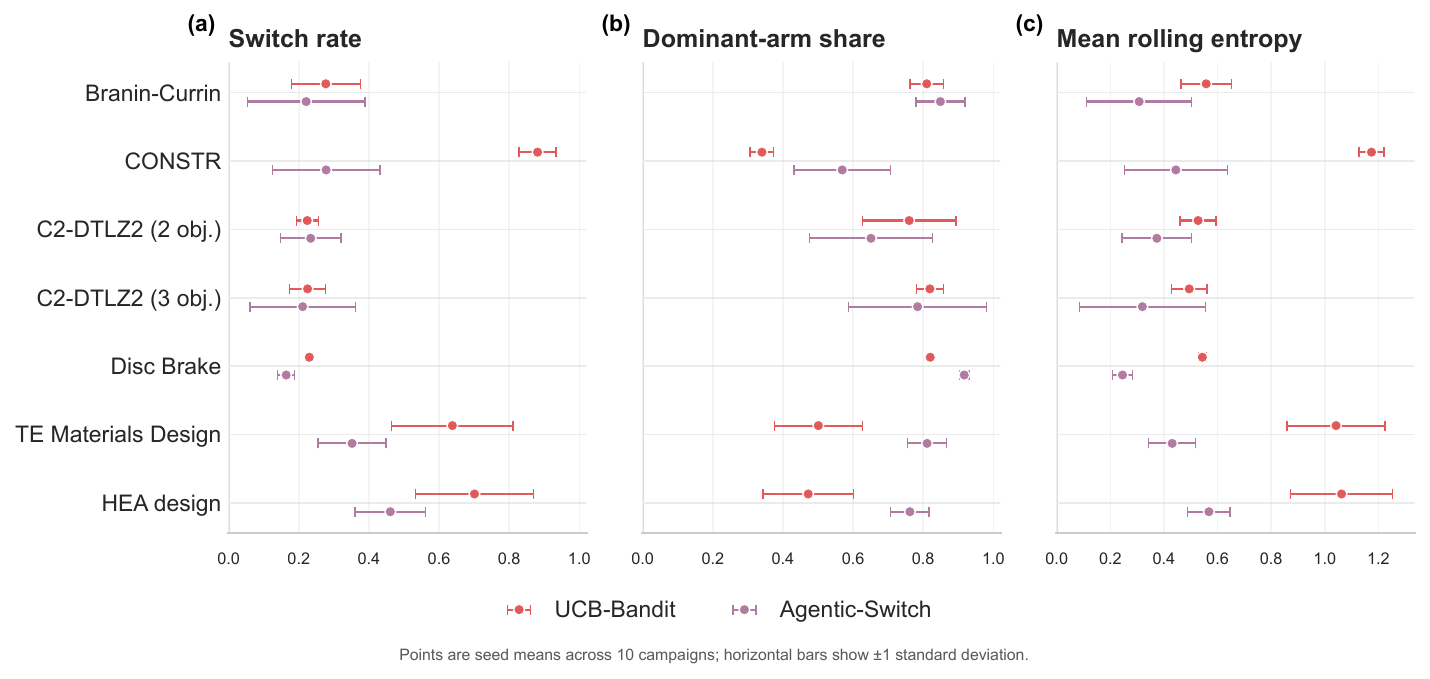}
    \caption{\textbf{Acquisition-switching diversity and persistence.}
    Adaptive-controller behavior is summarized by switch rate, dominant-arm share, and mean rolling entropy over 10-iteration windows. Points show seed means over 10 campaigns, and horizontal bars show \(\pm 1\) standard deviation. UCB-Bandit generally showed a higher switch rate and higher selection entropy, whereas Agentic-Switch was more persistent and more concentrated on selected acquisition families.}
    \label{fig:switching-diversity}
\end{figure*}

On CONSTR, \texttt{qLogPoF}, \texttt{qUCB}, and \texttt{qLogEHVI} drew comparable rewards, which accounts for the high switching rate in figure \ref{fig:switching-diversity}. Reward sparsity mattered most in the materials studies, where \texttt{qUCB} had only partial reward support in the thermoelectric problem and PoF was the sole arm with meaningful reward in the HEA problem. Non-feasibility selections there mostly reflected UCB-bonus sampling under sparse feedback, and further detail is given in section S4 of the supplementary material. Rolling entropy over a ten-iteration window supported the same reading, with UCB-Bandit maintaining higher acquisition-selection entropy than Agentic-Switch throughout. UCB-Bandit therefore reacted mainly to realized reward, which is what a non-contextual bandit controller does, whereas Agentic-Switch held to a selection over longer stretches once the state warranted it.


\section{Discussion}
\label{sec:discussion}
Across the synthetic benchmarks and the two materials design case studies, no fixed acquisition function performed well on both metrics. The direction of that trade-off was not consistent. On CONSTR, \texttt{qLogPoF} accumulated the most feasible points while finishing weakest in feasible hypervolume, and Disc Brake followed the same pattern. On both C2-DTLZ2 problems the ordering reversed, with \texttt{qLogNParEGO} reaching the highest feasible hypervolume from far fewer feasible points. The cost of choosing wrongly grew with the severity of the constraints. In the HEA study, where only 18 of 40,553 candidates were feasible and Pareto optimal, \texttt{pEHVI} and \texttt{qUCB} exhausted their budgets without returning a single feasible alloy. The behavior a problem requires is not known before a campaign begins, which makes a single up-front choice of acquisition function very risky. Both controllers instead made that choice repeatedly, conditioned on the state of the campaign. Across all seven problems they stayed close to the best fixed policy on both metrics, a consistency no single acquisition function achieved.

We studied a modified UCB-Bandit because it addresses the bandit problem in a mathematically interpretable way and offers a point of comparison with the less interpretable agentic controller. The comparison, however, is not perfectly symmetric. UCB-Bandit received a scalar reward combining feasible discovery and feasible-hypervolume gain under weights we set in advance, \(w_{\mathrm{feas}}=0.7\) and \(w_{\mathrm{HV}}=0.3\). That reward was deliberately feasibility-heavy, which is defensible for a risk-averse meta-policy, though the right balance is no easier to fix beforehand than the acquisition function itself. Our post-hoc analysis also showed the effective reward distribution shifting across the early, middle, and late stages of a campaign. Portfolio methods can inherit sublinear regret from UCB only when a UCB arm is present in the portfolio \cite{brochu2011portfolioallocationbayesianoptimization, sheikh2022bayesian}, and \texttt{qUCB} is present here. The obstacle is the reward signal itself, which is non-stationary and hand-weighted, so the classical finite-time guarantees \cite{auer2002finite} should not be assumed to characterize these experiments.

LLMs are trained on next-token prediction, yet strong performance on that objective appears to require internal structure that generalizes beyond it \cite{brown2020language, yao2022react}. Agentic-Switch produced distinct state-space trajectories across the seven problems, which is at least consistent with state-conditioned selection, although our experiments cannot establish what representation supports it. The three-agent design offers one reading of how a selection is reached. The feasibility advocate and the hypervolume advocate open two separate paths through the decision space, each reading the same optimization state through a different semantic lens. The arbiter then resolves the two recommendations against that state. Distinctions the agents draw in language, such as stalled progress against recent luck, are available to the arbiter at once. A bandit controller can shift only after realized reward has accumulated enough to move its estimates.

Several directions extend from this work. The acquisition portfolio can be treated as a design variable in its own right. We used \texttt{qUCB} at a fixed \(\beta = 0.2\), and raising that value would give the portfolio a more explicitly exploratory arm. Hybrid arms such as \texttt{EHVI}\(\times\)\texttt{PoF} or \texttt{qUCB}\(\times\)\texttt{PoF} could be added as well, each balancing feasibility against Pareto improvement or exploration within a single arm. We left those out here to keep the tension between feasibility seeking and feasible-hypervolume improvement visible in interpretable acquisition functions. The approach also extends to batch constrained MOBO, and contextual policies remain open, particularly under the low-data conditions typical of autonomous materials design campaigns.

\section{Methods}
\label{sec:methods}
\subsection{Constrained-MOBO Formulation}
\label{subsec:C-MOBO_formulation}
We considered constrained multi-objective optimization problems in which each candidate \(\x\in\X\) carries an objective vector \(\fvec(\x)=[f_1(\x),\ldots,f_M(\x)]^\top\in\R^M\) and a vector of constraint functions \(\cvec(\x)=[c_1(\x),\ldots,c_K(\x)]^\top\in\R^K\). All objectives were expressed under a common maximization convention, and all constraints were written as inequalities of the form \(c_k(\x)\leq0\). The optimization problem is then

\begin{equation}
    \label{eq:main-cmobo}
    \begin{aligned}
        \max_{\x \in \X} \quad & \fvec(\x) \\
        \text{subject to} \quad & c_k(\x) \leq 0,
        \quad k = 1,\ldots,K
    \end{aligned}
\end{equation}

The feasible set is 

\begin{equation}
    \label{eq:main-feasible-set}
    \F = \left\{\x\in\X: c_k(\x)\leq 0\ \forall k\right\}
\end{equation}

A candidate \(\x^\star\in\F\) is \textit{feasible non-dominated} if no other \(\x\in\F\) satisfies \(f_m(\x)\geq f_m(\x^\star)\) for all \(m\) with strict inequality for at least one objective. At each iteration \(t\), the feasible non-dominated set \(\Pset_t\subseteq\F\) was used to compute feasible hypervolume relative to a fixed reference point \(\mathbf{r}\). This was one of the two primary measures of optimization progress.

\subsection{Surrogate Models}

Following standard protocols for MOBO with unknown constraints, each objective \(f_m\) and each explicit constraint \(c_k\) was modeled with an independent single-output GP \cite{williams2006gaussian, gelbart2014bayesian}. Input features were normalized before model fitting, and outputs were standardized unless their variance was negligible. For objective \(m\) and constraint \(k\), the posterior predictive marginals given data \(\D_t\) are

\begin{equation}
\label{eq:main-gp-posteriors-1}
\begin{aligned}
f_m(\x)\mid\D_t\sim\mathcal{N}\!\left(\mu_{m,t}(\x),\sigma^2_{m,t}(\x)\right), \\
c_k(\x)\mid\D_t\sim\mathcal{N}\!\left(\mu_{c_k,t}(\x),\sigma^2_{c_k,t}(\x)\right).
\end{aligned}
\end{equation}

The predictive probability that a candidate $\mathbf{x}$ satisfied explicit constraint \(k\) is
\begin{equation}
\label{eq:main-constraint-prob}
p_{k,t}(\x)=\Pr\!\left[c_k(\x)\leq0\mid\D_t\right]
=\Phi\!\left(\frac{-\mu_{c_k,t}(\x)}{\sigma_{c_k,t}(\x)}\right),
\end{equation}
where \(\Phi\) is the standard normal cumulative distribution function. The joint probability of satisfying all explicit constraints was approximated under conditional independence across constraint models:
\begin{equation}
\label{eq:main-joint-pof}
p_{\mathrm{feas},t}(\x)\approx\prod_{k=1}^K p_{k,t}(\x).
\end{equation}

Although the constraints are unlikely to be independent, this approximation has proven effective in practice \cite{maguire2025good}. This probability-of-feasibility (PoF) quantity was used directly in the synthetic and thermoelectric finite-pool studies. The custom PoF used for the HEA case study is described in section \ref{subsec-methods-hea}.

\subsection{Acquisition portfolio}

Throughout this work, we used a common acquisition-function portfolio, with problem-specific modifications for the HEA case study. The portfolio was
\begin{equation}
\label{eq:main-arm-set}
\A=\left\{\texttt{qLogEHVI},\texttt{qLogNParEGO},\texttt{qUCB},\texttt{qLogPoF}\right\}.
\end{equation}

The portfolio was chosen to span complementary search behaviors: feasible-frontier improvement (\texttt{qLogEHVI}), scalarized trade-off exploration (\texttt{qLogNParEGO}), uncertainty-driven exploration (\texttt{qUCB}), and feasibility seeking (\texttt{qLogPoF}) \cite{balandat2020botorch}. The exploration parameter \(\beta\) for \texttt{qUCB} was fixed to 0.2. For the HEA case study we used custom acquisition functions adapted from Maguire et al.~\cite{maguire2025good}, substituting \texttt{pEHVI} for \texttt{qLogEHVI} and a custom \texttt{PoF} for \texttt{qLogPoF}.

At iteration \(t\), each acquisition arm \(a\in\A\) proposes a candidate \(\hat{\x}_{a,t}\). A policy \(\pi_t\) then selects which proposal is evaluated:
\begin{equation}
\label{eq:main-policy}
a_t=\pi_t(s_t),
\end{equation}
where \(s_t\) summarizes the current optimization state, including metadata, feasible-count history, hypervolume history, per-arm statistics, and recent outcomes.

\subsection{UCB-Bandit}

UCB-Bandit treated the acquisition functions in the portfolio as arms and maintained an empirical reward estimate for each, following classical UCB allocation \cite{auer2002finite}. Every arm was tried once during a warm start. Thereafter, the selected arm maximized
\begin{equation}
\label{eq:main-ucb-bandit}
\mathrm{UCB}_{j,t}=\hat{\mu}_{j,t}+c\sqrt{\frac{\log(t+1)}{\max\{1,n_{j,t}\}}},
\end{equation}
where \(\hat{\mu}_{j,t}\) is the empirical mean reward of arm \(j\), \(n_{j,t}\) is the number of previous selections of that arm, and \(c=1.0\) is the exploration constant.

The bandit reward combined feasible discovery and feasible-hypervolume gain:

\begin{equation}
\label{eq:main-bandit-reward}
\begin{aligned}
r_t
&= w_{\mathrm{feas}}\mathbf{1}[\Delta N_{\mathrm{feas},t}>0] \\
&\quad + w_{\mathrm{HV}}
\min\!\left(
    \frac{\max(\Delta \mathrm{HV}_t,0)}{\gamma_t},
    1
\right).
\end{aligned}
\end{equation}

with \(w_{\mathrm{feas}}=0.7\), \(w_{\mathrm{HV}}=0.3\), and \(\gamma_t\) equal to the running median of strictly positive hypervolume gains, defaulting to 1 when no positive gain had yet been observed.


\subsection{Agentic-Switch}

We used an LLM-based three-agent framework for adaptive switching among the acquisition functions in the portfolio. A feasibility advocate agent recommended the best arm for feasible discovery and feasibility protection, while a hypervolume advocate recommended the best arm for Pareto-front improvement. An arbiter agent then resolved the two recommendations against the current state. Agentic-Switch factorized the policy \(\pi_t\) in \cref{eq:main-policy} as


\begin{figure*}[tbh!]
\centering
\begin{tikzpicture}[
    x=1cm,y=1cm,
    >={Latex[length=2.8mm,width=2mm]},
    font=\sffamily\small,
    flow/.style={
        draw=black!80,
        line width=0.95pt,
        -{Latex[length=2.8mm,width=2mm]}
    },
    box/.style={
        rounded corners=2.5pt,
        draw=black!75,
        line width=0.9pt,
        fill=white,
        align=center,
        inner sep=6pt
    },
    statebox/.style={
        box,
        fill=black!03,
        text width=3.1cm,
        minimum height=1.7cm
    },
    agentbox/.style={
        box,
        text width=3.15cm,
        minimum height=1.35cm
    },
    arbiterbox/.style={
        box,
        fill=black!05,
        text width=2.9cm,
        minimum height=1.55cm
    },
    outputbox/.style={
        box,
        draw=black!90,
        fill=black!90,
        text=white,
        text width=3.15cm,
        minimum height=1.15cm
    },
    groupbox/.style={
        draw=black!50,
        dashed,
        dash pattern=on 4pt off 3pt,
        rounded corners=3pt,
        line width=0.85pt
    },
    grouplabel/.style={
        font=\sffamily\scriptsize\bfseries,
        text=black!65,
        fill=white,
        inner sep=1.5pt
    }
]
\node[statebox] (state) at (0,0) {%
\textbf{State packet}\\[2pt]
\footnotesize
global progress\\
arm statistics\\
recent history
};
\node[agentbox] (feas) at (4.7,1.1) {%
\textbf{Feasibility Advocate}\\[2pt]
\footnotesize
protect feasible\\
discovery
};
\node[agentbox] (hv) at (4.7,-1.1) {%
\textbf{HV Advocate}\\[2pt]
\footnotesize
promote feasible-frontier\\
growth
};
\node[arbiterbox] (arbiter) at (9.2,0) {%
\textbf{Arbiter}\\[2pt]
\footnotesize
resolve\\
recommendations
};
\node[outputbox] (output) at (13.1,0) {%
\textbf{Final arm}\\[2pt]
\footnotesize
selection
};
\node[groupbox, fit=(feas)(hv), inner xsep=0.45cm, inner ysep=0.38cm] (group) {};
\node[grouplabel, above=2pt of group.north] {Parallel LLM specialist review};

\draw[flow] (state.east) -- ++(0.55,0) |- (feas.west);
\draw[flow] (state.east) -- ++(0.55,0) |- (hv.west);

\draw[flow] (feas.east) -- ++(0.55,0) |- ([yshift=2pt]arbiter.west);
\draw[flow] (hv.east) -- ++(0.55,0) |- ([yshift=-2pt]arbiter.west);

\draw[flow] (arbiter.east) -- (output.west);

\end{tikzpicture}
\caption{Schematic representing the agentic framework used in this work for adaptive acquisition policy switching. A structured Bayesian optimization state is provided to two parallel specialist agents, one focused on feasible discovery and one focused on feasible-frontier growth. Their recommendations are then resolved by an arbiter to select the acquisition arm used at iteration $t$.}
\label{fig:llm-switch-workflow}
\end{figure*}
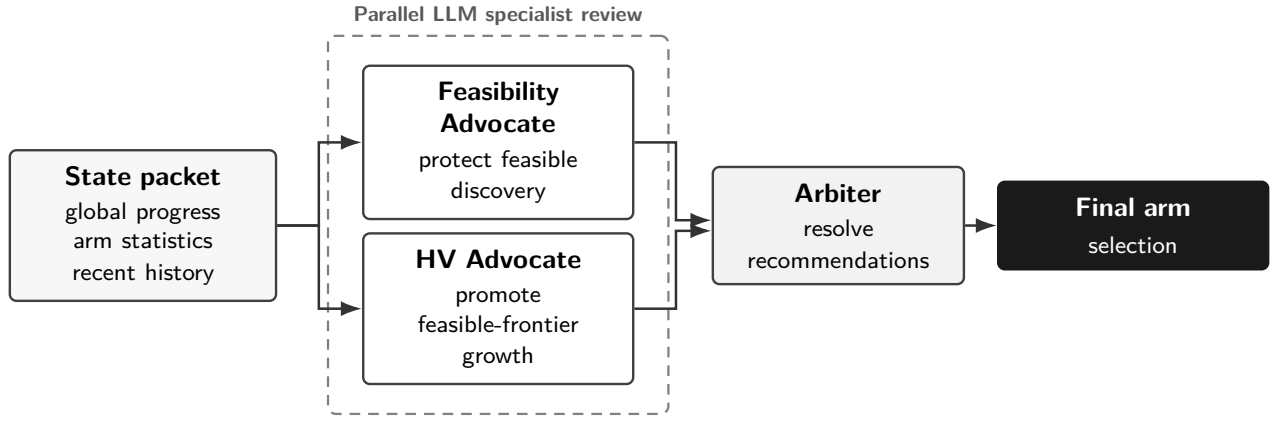

\begin{equation}
    \label{eq:main-llm-factorization}
    \begin{aligned}
    \tilde{a}^{(\mathrm{feas})}_t
    &= \psi^{(\mathrm{feas})}_t(s_t), \\
    \tilde{a}^{(\mathrm{HV})}_t
    &= \psi^{(\mathrm{HV})}_t(s_t), \\
    a_t
    &= \psi^{(\mathrm{arb})}_t\!\left(
        s_t,
        \tilde{a}^{(\mathrm{feas})}_t,
        \tilde{a}^{(\mathrm{HV})}_t
    \right).
    \end{aligned}
\end{equation}

where \(\tilde{a}^{(\mathrm{feas})}_t\) and \(\tilde{a}^{(\mathrm{HV})}_t\) are the advocate recommendations and \(a_t\in\A\) is the final selected arm. Agentic-Switch used the same action space \(\A\) as UCB-Bandit. At any iteration \(t\), the transmitted state packet was

\begin{equation}
    \label{eq:main-llm-state}
    s_t=\left(m_t,\ g_t,\ \{d_{a,t}\}_{a\in\A},\ h_t\right),
\end{equation}
where \(m_t\) denotes metadata, \(g_t\) the global optimization statistics, \(d_{a,t}\) the current statistics for arm \(a\), and \(h_t\) a memory window of recent iterations. The metadata and calculated statistics are detailed in Supplementary Section S1. Each advocate agent returned a strict JSON object containing an acquisition recommendation \(\tilde{a}_t\in\A\), a short snake-case reason code, a natural-language argument, and a confidence score. The arbiter returned a strict JSON object containing the final acquisition label \(a_t\in\A\), a short snake-case reason code, a reflection resolving the two advocate recommendations, and a confidence score. Only \(a_t\) was passed to the optimizer. The accompanying text was logged for audit and human interpretation.

\subsection{Synthetic Benchmarks}

We evaluated our methodology on general-purpose constrained MOBO through five synthetic benchmark problems: constrained Branin--Currin ($d=2$, $m=2$, $c=1$), CONSTR ($d=2$, $m=2$, $c=2$), two- and three-objective C2-DTLZ2 variants ($d=11$, $m=2$, $c=1$; $d=12$, $m=3$, $c=1$), and Disc Brake ($d=4$, $m=2$, $c=4$). Each problem was run for 100 iterations across 10 seeds, each initialized with a scrambled Sobol design.

\subsection{Thermoelectric Materials Design}

The thermoelectric case study was formulated as a finite-pool constrained MOBO using data from Na et al.~\cite{na2022public}. The dataset was preprocessed to obtain a pool of 3,138 candidates in the mid-temperature region (\(300\leq T\leq600\ \mathrm{K}\)) with 856 unique formulas, 144 feasible candidates, and 82 feasible Pareto candidates. Each candidate was represented by a 13-dimensional feature vector containing chemical descriptors and normalized temperature. We maximized the absolute Seebeck coefficient (\(|S|\)) and the electrical conductivity (\(\log_{10}\sigma\)), and minimized the thermal conductivity (\(\kappa\)). The constraints were set on the thermoelectric figure of merit (\(ZT\geq0.8\)) and the power factor (\(\mathrm{PF}\geq1.2\times10^{-3}\ \mathrm{W\,m^{-1}\,K^{-2}}\)), where

\begin{equation}
\label{eq:main-estm-pf-zt}
\begin{aligned}
\mathrm{PF}
&= S^2 \sigma, \\
ZT
&= \frac{S^2 \sigma T}{\kappa}
 = \frac{\mathrm{PF}\,T}{\kappa}.
\end{aligned}
\end{equation}

Both constrained properties are deterministic functions of the three objectives and the measurement temperature. We fit a separate GP to each of them, so that the constraint surrogates stayed calibrated against the tabulated values instead of inheriting error propagated through \cref{eq:main-estm-pf-zt}. Each run drew \(n_{\mathrm{init}}=8\) initial candidates uniformly at random from the pool, followed by 100 optimization iterations over 10 paired seeds.

\subsection{High Entropy Alloy Design}
\label{subsec-methods-hea}
The high-entropy alloy (HEA) case study was formulated as a finite-pool constrained MOBO with an additional hidden staged feasibility constraint. The data and problem formulation were taken directly from Maguire et al.~\cite{maguire2025good}. The design space was the Nb-Mo-Ta-V-W-Cr system, with 40,553 alloys in the pool. Of these, 12,591 (\(31.05\%\)) were Pareto-optimal, 22 (\(0.054\%\)) met the constraints, and only 18 (\(0.044\%\)) were both Pareto-optimal and feasible. We maximized solidus temperature (\(ST\)), Pugh ratio (\(P\)), and yield strength (\(YS\)), and minimized density (\(\rho\)). The constraints were of two kinds: objective-based constraints (\(ST>2473~\mathrm{K}\), \(\rho<9.0~\mathrm{g\,cm^{-3}}\), \(YS>700~\mathrm{MPa}\), and \(P>2.5\)), and a manufacturing-feasibility constraint based on phase stability (\(x_{\mathrm{BCC}}\geq 99\%\) at \(600^{\circ}\mathrm{C}\)).

Every queried alloy produced a phase observation, but the objectives were observed only if the alloy passed the BCC phase-stability constraint. GPs for the objectives and their constraints were fit only to the fully measured alloys, whereas the BCC-screening surrogate was fit to all phase-observed alloys.

Let \(p_{\mathrm{BCC},t}(\x)\) denote the predicted probability that candidate \(\x\) passes the BCC screen, obtained by mapping the posterior mean of the binary BCC indicator through a logistic sigmoid. Downstream property-pass probabilities were computed from the posterior distributions of \(ST\), \(\rho\), \(YS\), and \(P\), and their product defined \(p_{\mathrm{obj},t}(\x)\). The total predicted feasibility was
\begin{equation}
\label{eq:main-hea-total-feas}
p_{\mathrm{total},t}(\x)=p_{\mathrm{BCC},t}(\x)p_{\mathrm{obj},t}(\x).
\end{equation}
 
The acquisition scores used the same action space \(\A\) but were weighted by the probability of passing the BCC screen. If \(a^{\mathrm{base}}_{j,t}(\x)\) is the downstream score from arm \(j\), the HEA score was
\begin{equation}
\label{eq:main-hea-bcc-weighting}
a^{\mathrm{HEA}}_{j,t}(\x)=a^{\mathrm{base}}_{j,t}(\x)p_{\mathrm{BCC},t}(\x).
\end{equation}

Each run drew a single initial alloy uniformly at random from the pool (\(n_{\mathrm{init}}=1\)) and observed it under the same staged protocol as every later query, followed by 100 optimization iterations over 10 paired seeds.

\subsection{Evaluation Metrics}

We used feasible hypervolume and cumulative feasible count as the two primary performance metrics. Feasible hypervolume was computed from the feasible non-dominated candidates under the common maximization convention adopted above. Given a feasible non-dominated set \(\Pset\) and a reference point \(\mathbf{r}\), feasible hypervolume was defined as

\begin{equation}
\label{eq:main-hv}
\begin{aligned}
\HV(\Pset;\mathbf{r})
&=
\lambda\left(
\bigcup_{\x\in\Pset}
[r_1,f_1(\x)]
\right. \\
&\qquad\left.
{}\times \cdots \times
[r_M,f_M(\x)]
\right).
\end{aligned}
\end{equation}

where \(\lambda\) is the Lebesgue measure on \(\R^M\). For the adaptive controllers we also evaluated the switch rate, defined as the fraction of iterations at which the selected arm differed from the arm selected at the previous iteration. Rolling entropy was calculated over a moving 10-iteration window:

\begin{equation}
    \label{eq:main-entropy}
    \begin{aligned}
        H_t = - \sum_a p_t(a) \log p_t(a),
    \end{aligned}
\end{equation}
where \(p_t(a)\) is the fraction of iterations in the window at which arm \(a\) was selected. We used \(H_t\) to compare switching behavior between the two adaptive controllers.

\subsection{Implementation details}
Gaussian process surrogate models and acquisition functions were implemented with BoTorch \cite{balandat2020botorch}, which uses GPyTorch \cite{gardner2018gpytorch} for scalable GP inference. All experiments were run sequentially with one selected candidate per iteration (\(q=1\)). The agentic framework was built on a LangGraph-backed JSON client \cite{langgraph2026}. The two advocates and the arbiter all used \texttt{gpt-4o} \cite{openai2024gpt4o} through the OpenAI provider, with temperature \(0.1\), a timeout of \(90\) s, and a memory window of \(10\) recent iterations. Each agent received the same structured optimization state under a role-specific prompt and was required to return strict JSON. All computations were performed on an NVIDIA DGX Spark system.

\section*{Data availability}
The data supporting the results of this study are available at https://github.com/sushant-k-sinha/PCMOBO.

\section*{Code availability}
The code supporting the results of this study are available at https://github.com/sushant-k-sinha/PCMOBO

\section*{Acknowledgments}
The research was sponsored by the Army Research Laboratory and was accomplished under Cooperative Agreement Number W911NF-22-2-0106. The views and conclusions contained in this document are those of the authors. They should not be interpreted as representing the official policies, either expressed or implied, of the Army Research Laboratory or the US Government. The US Government is authorized to reproduce and distribute reprints for Government Purposes, notwithstanding any copyright notation herein. Original data were generated within the BIRDSHOT Center (https://birdshot.tamu.edu), supported by the Army Research Laboratory under Cooperative Agreement (CA) Number W911NF-22-2-0106 (All Authors acknowledge support from this CA).

\section*{Author contributions}
S.S. developed the algorithms, performed the experiments, and analyzed results. C.H. wrote the BO code for HEA problem. S.S., R.R. and S.P.P. ideated the study. R.A., D.A., S.P.P. and B.V. supervised the research. R.A. secured funding for the work. All the authors contributed in preparing the manuscript.

\section*{Competing interests}
The authors declare no competing interests.


\includepdf[pages=-]{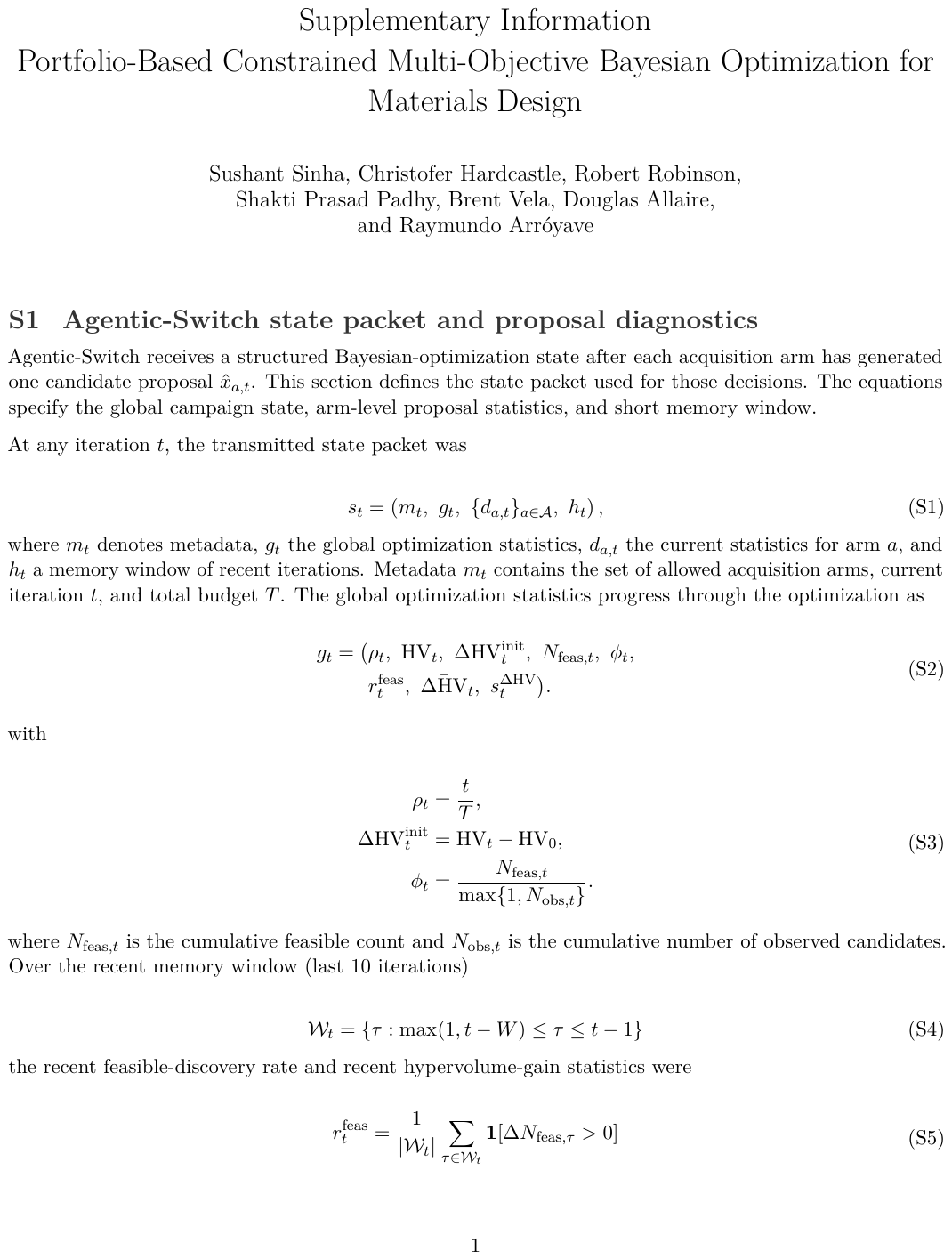}

\end{document}